\documentstyle[epsfig,cite]{mn2e}
\begin{document}

\title[The 2003 radio outburst of a new X-ray transient: XTE~J1720$-$318]
{The 2003 radio outburst of a new X-ray transient: XTE~J1720$-$318}
\author[Brocksopp et al.]
{C.~Brocksopp$^1$\thanks{email: cb4@mssl.ucl.ac.uk}, S.~Corbel$^2$, R.P.~Fender$^3$, M.~Rupen$^4$, R.~Sault$^5$, S.J.~Tingay$^{6}$,
\newauthor D.~Hannikainen$^7$, K.~O'Brien$^8$\\
$^1$Mullard Space Science Laboratory, Holmbury St. Mary, Dorking, Surrey RH5 6NT\\
$^2$Universit\'e Paris 7 Denis Diderot and Service d'Astrophysique, UMR AIM, 
CEA Saclay, F-91191 Gif sur Yvette, France.\\
$^3$Astronomical Institute ``Anton Pannekoek'', University of Amsterdam and Center for High Energy Astrophysics,\\
Kruislaan 403, 1098 SJ Amsterdam, The Netherlands\\
$^4$National Radio Astronomy Observatory, Socorro, NM 87801, USA\\
$^5$Australia Telescope National Facility, P.O. Box 76, Epping, NSW 1710, Australia\\
$^6$Centre for Astrophysics and Supercomputing, Swinburne University of Technology, Mail Number 31, P.O. Box 218,\\ Hawthorn, VIC 3122, Australia\\
$^7$Observatory, PO Box 14, FIN-00014 University of Helsinki, Finland\\
$^8$European Southern Observatory, Casilla 19001, Santiago 19, Chile\\
}

\date{Accepted ??. Received ??}
\pagerange{\pageref{firstpage}--\pageref{lastpage}}
\pubyear{??}
\maketitle

\begin{abstract}
We present radio observations of the black hole X-ray transient XTE J1720$-$318, which was discovered in 2003 January as it entered an outburst. We analyse the radio data in the context of the X-ray outburst and the broad-band spectrum. An unresolved radio source was detected during the rising phase, reaching a peak of nearly 5 mJy approximately coincident with the peak of the X-ray lightcurve. Study of the spectral indices suggests that at least two ejection events took place, the radio-emitting material expanding and becoming optically thin as it faded. The broad-band spectra suggested that the accretion disc dominated the emission, as expected for a source in the high/soft state. The radio emission decayed to below the sensitivity of the telescopes for $\sim 6$ weeks but switched on again during the transition of the X-ray source to the low/hard state. At least one ``glitch'' was superimposed on the otherwise exponential decay of the X-ray lightcurve, which was reminiscent of the multiple jet ejections of XTE J1859+226. We also present a $K_s$-band image of XTE J1720$-$318 and its surrounding field taken with the VLT.
\end{abstract}


\section{Introduction}

\begin{figure}
\begin{center}
\psfig{file=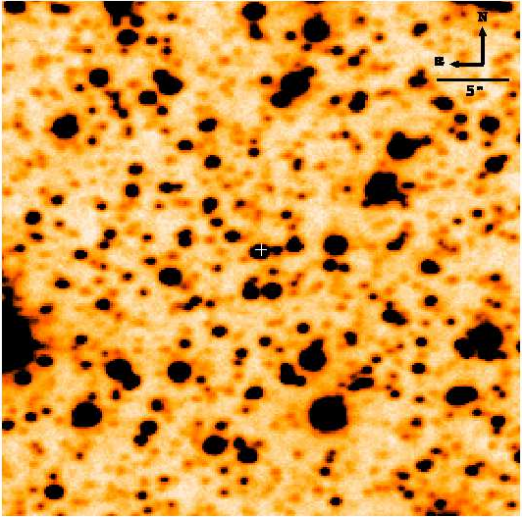,angle=0,width=7cm}
\vspace*{0cm}
\caption{VLT/ISSAC image of the field of XTE J1720$-$318 in the $K_s$-band. The position of the infrared source is marked with a cross.}
\label{ir-image}
\end{center}
\end{figure}

X-ray transient sources are X-ray binary systems, typically with a low mass companion star and a black hole for the compact object (although sometimes a neutron star). They are well-known for dramatic outbursts caused by some form of instability within the accretion flow. There are now $\sim 40$ observed sources, some of which are recurrent although the majority have still only been observed in outburst once.

The canonical ``soft'' X-ray transient outburst is one which displays a soft blackbody spectrum in low energy X-rays and a ``Fast Rise Exponential Decay'' lightcurve morphology (e.g. Chen, Shrader \& Livio 1997). With modern X-ray telescopes it has been possible to obtain the spectral and temporal coverage to show that the behaviour is more complicated than this. X-ray transients appear to enter the outburst from an initially hard spectral state (the low/hard state; see e.g. van der Klis 1995). On a timescale of $\sim$ days--weeks the X-ray source then softens in most cases (e.g. Brocksopp et al. 2002). Some X-ray transients, however, do not soften but remain in the low/hard state throughout the outburst (e.g. Brocksopp et al. 2004, 2001).

All black hole X-ray transient sources which have been observed at radio frequencies have been detected at some point during their outburst\footnote{The one exception to this is XTE J1755$-$324 but only a single upper limit was obtained and the observation took place during the decay, $\sim 1$ month after the onset of the outburst (Ogley et al. 1997)} (Brocksopp et al. 2002), although the relationship between the X-ray and radio lightcurves is not a simple correlation. In most of these a jet origin can be inferred, either from synthesis imaging or from study of the spectral index of the radio emission. Low/hard X-ray states are associated with a flat-spectrum compact jet whereas the high/soft (now also known as thermal-dominant) state is not (Fender 2003, 2001; McClintock \& Remillard 2003). Short-lived discrete ejections are also observed and appear to take place during the transition from the hard to the very high state (or steep power-law state). In particular XTE J1859+226 was observed to make a series of 5--6 radio ejections, each of which was associated with a temporary epoch of X-ray hardening superimposed on a general softening during the decay (Brocksopp et al. 2002).

\subsection{XTE J1720$-$318}

XTE J1720$-$318 was detected for the first time in 2003 January by the All Sky Monitor on-board the Rossi Timing X-ray Explorer ({\sl RXTE}/ASM; Remillard et al. 2003). The average 2--12 keV flux was $\sim130$ mCrab initially, later rising to $\sim430$ mCrab. The source was hard at first but was later observed by the {\sl RXTE}/PCA to have a soft spectrum and a low level of high-frequency variability, consistent with the high/soft state (Markwardt \& Swank 2003; see also e.g. van der Klis 1995). A later {\sl RXTE}/PCA observation revealed an iron line at 6.2 keV with a FWHM of 2.5 keV and equivalent width of 95 eV (Markwardt 2003). 

The radio counterpart was discovered by Rupen et al. (2003) and was found to display variability on a timescale of $\sim$days; the flux density peaked at 4.9 mJy at 4.9 GHz. The infrared counterpart was discovered by Kato et al. (2003) with $J$, $H$ and $K_s$ band magnitudes in the range 15--17. O'Brien et al. (2003) confirmed the $K_s$ band detection with VLT observations and also refined the radio position of the source to RA=17:19:58.994, DEC=$-$31:45:01.25. Further infared monitoring by Nagata et al. (2003) revealed a decaying lightcurve, superimposed by a secondary maximum $\sim40$ days after the outburst; the infrared light was interpreted as emission from an irradiated accretion disc.

Finally, XTE J1720$-$318 was observed towards the end of its decay from outburst by {\sl INTEGRAL} (during pointed observations of H1743$-$322) and was found to be in the low/hard state (Cadolle Bel et al. 2004).

\section{Observations}

\subsection{Radio}

\begin{table*}
\caption{Observation dates, integration times, flux densities and $1\sigma$ errors for the ATCA (A) and VLA (V) observations. Non-detections are shown by $3\sigma$ upper limits. The final column indicates whether the X-ray source was in the low/hard or a soft (i.e. high/soft, intermediate or very high) state at the time of the radio observation.}
\center
\label{observations}
\begin{tabular}{lcccccc}
\hline
UT Date & MJD&Integration Time & $S_{1.4 GHz}$& $S_{4.8,\, 4.9\, GHz}$&$S_{8.5,\, 8.6\, GHz}$& X-ray\\ 
&&(Hours) + Telescope & (mJy) & (mJy)&  (mJy)&State\\
\hline
2003 01 15& 52654.61& 0.49 (V) & $<0.71$ & --			& --		&soft\\
2003 01 15& 52654.62& 0.70 (V) & --       &$0.33\pm0.04$	& --		&soft\\  	 
2003 01 15& 52654.63& 0.31 (V) & --       & -- 			&$0.27\pm0.06$	&soft\\   	 
2003 01 16& 52655.80& 2.97 (A) & --       &$4.73\pm0.06$	& --    	&soft\\		
2003 01 16& 52655.84& 0.21 (V) & --       &$4.42\pm0.10$	& --		&soft\\   	 
2003 01 17& 52656.20& 1.22 (A) & --       &$1.74\pm0.15$	&$1.19\pm0.12$	&soft\\	
2003 01 18& 52657.79& 3.40 (A) & --       &$1.12\pm0.06$	&$0.91\pm0.05$	&soft\\	
2003 01 18& 52657.83& 0.10 (V) & --       &$1.26\pm0.11$	&$1.00\pm0.09$	&soft\\    	
2003 01 20& 52659.84& 1.37 (A) & --       &$1.30\pm0.05$	&$0.73\pm0.07$	&soft\\	
2003 01 21& 52660.76& 0.13 (V) & --       &$0.44\pm0.08$	& --		&soft\\    	
2003 01 21& 52660.77& 0.12 (V) & --       & --			&$0.39\pm0.07$	&soft\\    	
2003 01 29& 52668.75& 0.89 (A) & --       &$<0.38$        	&$<0.39$      	&soft\\	
2003 02 06& 52676.58& 0.25 (V) & --       &$<0.21$		& --		&soft\\   	 
2003 02 13& 52683.63& 0.22 (V) & --       & --			&$<0.18$	&soft\\   	 
2003 02 20& 52690.59& 0.22 (V) & --       & --			&$<0.54$	&soft\\   	 
2003 02 26& 52696.51& 0.29 (V) & --       &$<0.27$		& --		&soft\\   	 
2003 03 05& 52703.49& 0.29 (V) & --       &$<0.24$		& --		&soft\\   	 
2003 03 07& 52705.62& 0.28 (V) & --       &$<0.21$		& --		&soft\\   	 
2003 03 14& 52712.68& 0.20 (V) & --       &$<0.33$	        & --		&soft\\   	 
2003 03 17& 52715.47& 0.25 (V) & --       &$<0.21$		& --		&soft\\   	 
2003 03 30& 52728.58& 0.26 (V) & --       &$0.34\pm0.08$	& --		&low/hard\\   	 
2003 04 06& 52735.43& 0.25 (V) & --       &$0.23\pm0.10$	& --		&low/hard\\   	 
2003 04 08& 52737.49& 0.24 (V) & --       &$<0.21$		& --		&low/hard\\   	 
2003 04 16& 52745.41& 0.15 (V) & --       & --			&$<0.24	$	&low/hard\\    	
2003 04 25& 52754.57& 2.98 (A) & --       &$<0.22$        	&$<0.27$      	&low/hard\\	
2003 04 26& 52755.53& 0.42 (V) & --       & --			&$0.36\pm0.05$	&low/hard\\   	 
2003 04 26& 52755.54& 0.43 (V) & --       &$0.41\pm0.07$	& --		&low/hard\\   	
2003 04 27& 52756.73& 0.71 (A) & --       &$<0.42$        	&$<0.45$      	&low/hard\\	
2003 04 28& 52757.69& 1.90 (A) & --       &$<0.30$        	&$<0.30$      	&low/hard\\	
2003 05 06& 52765.42& 0.27 (V) & --       & --			&$<0.15$   	&low/hard or off?\\   	 
2003 05 20& 52779.45& 0.30 (V) & --       & --			&$<0.21$   	&low/hard or off?\\   	 
2003 07 09& 52829.34& 0.19 (V) & --       & --			&$<0.18$   	&low/hard or off ?\\   	 
2003 08 19& 52870.06& 0.29 (V) & --       &$<0.15$		& --		&low/hard or off?\\   	 
2003 08 25& 52877.01& 0.13 (V) & --       &$<0.24$		& --		&low/hard or off?\\    	
\hline
\end{tabular}
\end{table*}

The Australia Telescope Compact Array (ATCA) observations of XTE J1720$-$318 took place in 2003 January and April when the array was in the 6B and EW352 configurations respectively. The source was observed at 4.8 GHz and 8.6 GHz with a bandwidth of 128 MHz. The flux calibrators were PKS 1934$-$638 and PKS 0823$-$500 and the phase calibrator for all epochs was PMN J1733$-$3722. The data were reduced in the standard way with (minimal) flagging, flux and phase calibration and finally mapping using {\sc miriad}. A point source was fitted to the detected emission and the flux density measured. 

Additional radio observations at 4.9 and 8.5 GHz were obtained by the Very Large Array (VLA) and gave fluxes consistent with the ATCA data. Observations were made over an $\sim 8$ month period, during which time the array was in a number of different configurations; all detections took place while the array was in the DnC or D configurations. Two different phase calibrators were used, IERS B1741-312 for the 4-cm observations and PKS J1700-2610 for 6-cm. The data were reduced in the standard way using {\sc AIPS}.

A variable radio source was detected within the X-ray error circle with position RA=17:19:58.994,\\ Dec=$-$31:45:01.25 ($\pm 0.25\arcsec$). Further details of the observations and the resultant flux densities are listed in Table~\ref{observations}.

\subsection{Infrared}

We have also obtained a high-resolution infrared ($K_s$-band) image of the field of XTE J1720$-$318 using the ISAAC instrument on the Very Large Telescope. The observation took place on 2003 January 21 and we show the resultant image in Fig.~\ref{ir-image}. North is up, east is to the left and the image has a limiting magnitude of $\sim 20.2$ and a full width half maximum of $0.5\arcsec$. The source was detected with a $K_s$-band magnitude of $15.31\pm0.03$, consistent with the results of Nagata et al (2004). These authors obtained further monitoring and confirmed that the source was indeed the infrared counterpart to XTE J1720$-$318. There was no significant variability at the level of $\pm 0.1$ magnitudes during the 30-minute sequence of 60-second exposures. The infared source coinciding with the best radio position is marked with a cross.

\begin{figure*}
\begin{center}
\vspace*{-1cm}
\psfig{file=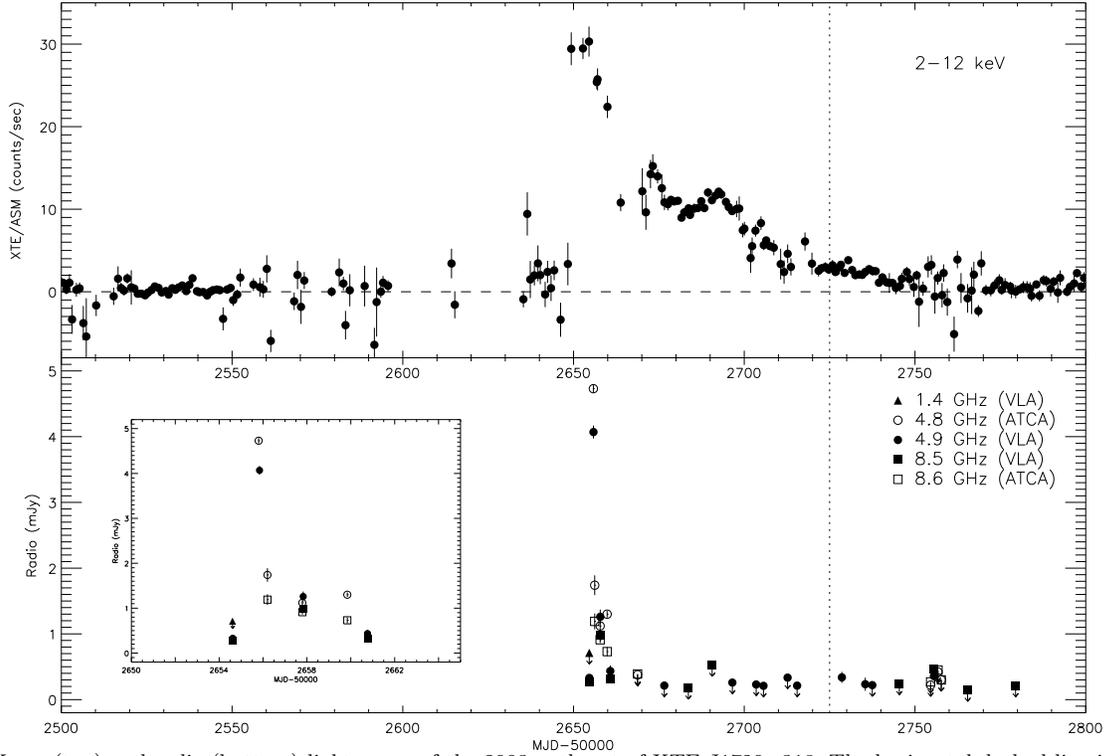,angle=0,width=16cm}
\vspace*{-1cm}
\caption{X-ray (top) and radio (bottom) lightcurves of the 2003 outburst of XTE J1720$-$318. The horizontal dashed line indicates an ASM count rate of zero. The vertical dotted line indicates the time at which {\sl RXTE}/PCA detected XTE J1720$-$318 in the low/hard state. The inset panel shows the radio lightcurve with an expanded time axis.}
\label{lightcurve}
\end{center}
\end{figure*}

\begin{figure*}
\begin{center}
\vspace*{-1cm}
\psfig{file=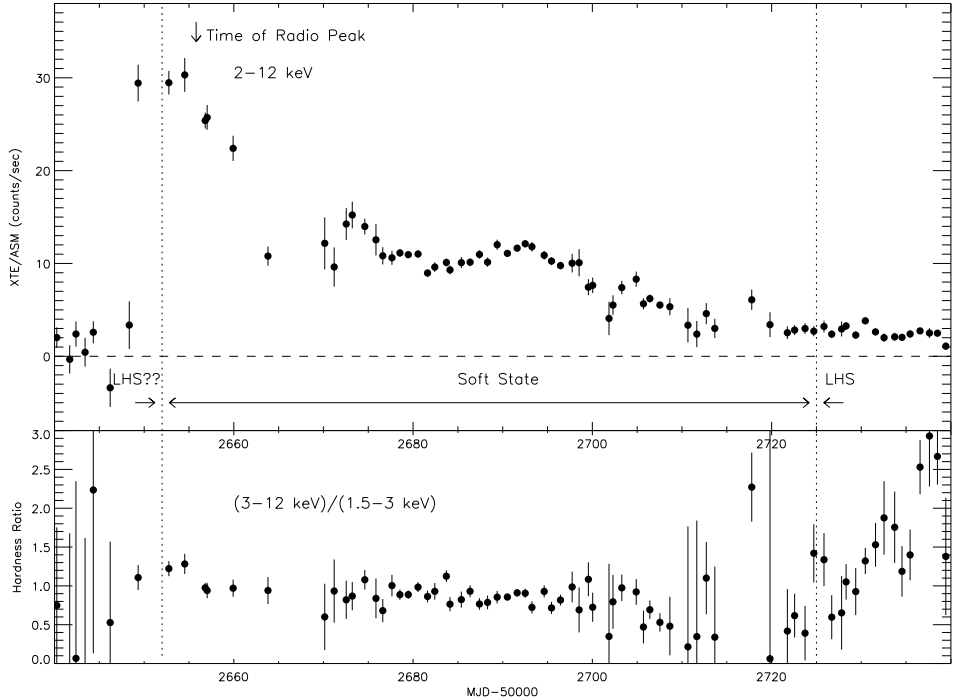,angle=90,width=14cm}
\vspace*{-0.5cm}
\caption{X-ray lightcurve (top) and hardness ratio (bottom) of the 2003 outburst of XTE J1720$-$318. The horizontal dashed line indicates an ASM count rate of zero. The vertical dotted lines indicate the approximate date of the transition from the low/hard state to a softer state (according to Remillard et al. 2003) and the date of the transition back to the low/hard state (J. Swank, private communication). }
\label{hardness}
\end{center}
\end{figure*}

\section{Results -- radio and X-ray lightcurves}

The radio counterpart to XTE J1720$-$318 was detected during 2003 January as an unresolved (the FWHM of the ATCA beam was $5.25\times 1.44 \arcsec $) variable source with both the ATCA and the VLA. Its flux density was initially at a low value ($<0.7$ mJy) but then rose to nearly 5 mJy within a day, peaking on 2003 January 16 (MJD 52655). The radio source was present for the following $\sim$ week before decaying to below the detection threshold of the telescopes. It was not detected significantly again until 2003 March 30. The radio fluxes and upper limits are listed in Table~\ref{observations} and shown in the bottom panel of Fig.~\ref{lightcurve}.

The top panel of Fig.~\ref{lightcurve} shows the 2-12 keV lightcurve obtained by {\sl RXTE}/ASM. Each data-point is the daily-averaged count rate. The X-ray source rose rapidly above quiescence in 2003 January, detected for the first time on January 9 (MJD 52648) when it was already close to its peak flux. The source was relatively hard during the initial observations but had reached the high/soft state by January 13 (MJD 52652; Remillard et al. 2003). The X-ray lightcurve reached a maximum on January 15 (MJD 52654), simultaneously with or possibly {\em prior} (as for XTE~J1650$-$500, Corbel et al. 2004) to the radio maximum but the time resolution of the radio observations is insufficient to be sure of this. 

The X-ray source made the transition to the low/hard state on MJD 52725--6 (as observed by {\sl INTEGRAL} and confirmed by {\sl RXTE}; Cadolle Bel et al. (2004), J. Swank, priv. comm.); this is indicated in Fig.~\ref{lightcurve} by the vertical dotted line. The low/hard state is typically associated with the presence of radio emission and, as expected, the radio counterpart to XTE J1720$-$318 was indeed detected at this time. 

After January 15 the X-ray source started to fade with at least one, and possibly a number of, secondary outbursts or ``glitches'' (using the terminology of Chen, Shrader \& Livio 1997) during the decay. In Fig.~\ref{hardness} we plot the hardness ratio (HR1: 3--12 keV/1.5--3 keV) for the duration of the outburst. There is little variability and the signal-to-noise ratio is poor but there is a number of interesting features. The source is hard at the onset of the outburst and then softens during the {\em decline} from the X-ray peak. Furthermore the radio peak takes place {\em after} the source has started to soften. During the softer period there is some variability but, due to the S/N, it is not clear whether or not there is a correlation between glitches and temporary hardening events. However, analysis of {\sl XMM-Newton} and {\sl RXTE}/PCA data by Cadolle Bel et al. (2004) suggests that there is little evidence for such short-term spectral variability. Following the transition to the low/hard state there is a significant increase in hardness.

\begin{table}
\caption{Spectral indices for the simultaneous 4.8--8.6 (ATCA) GHz and 4.9--8.5 (VLA) GHz data.}
\center
\label{alpha}
\begin{tabular}{lcccc}
\hline
UT Date & MJD & Telescope& $\alpha$ & $\alpha_{err}$\\
\hline
2003 01 15& 52654.6 &VLA & -0.36 & 0.17\\
2003 01 17& 52656.2 &ATCA & -0.65 & 0.11\\
2003 01 18& 52657.8 &ATCA & -0.36 & 0.06\\
2003 01 18& 52657.8 &VLA & -0.42 & 0.09\\
2003 01 20& 52659.8 &ATCA & -0.99 & 0.11\\
2003 01 21& 52660.8 &VLA & -0.22 & 0.16\\
2003 04 26& 52755.5 &VLA & -0.24 & 0.14\\
\hline		      
\end{tabular}
\end{table}

\begin{figure}
\begin{center}
\vspace*{-1cm}
\psfig{file=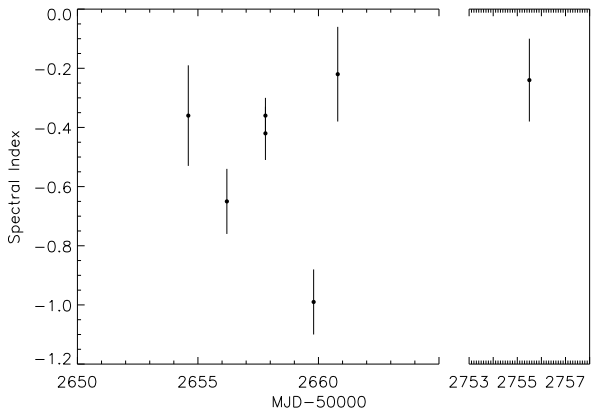,angle=90,width=8cm}
\caption{Spectral index plotted against time for those epochs for which simultaneous data at two frequencies was obtained. Note the break in the time axis.}
\label{alpha-plot}
\end{center}
\end{figure}

\begin{figure}
\begin{center}
\psfig{file=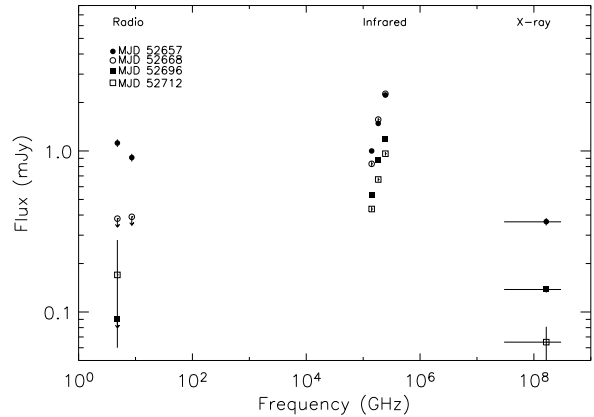,angle=90,width=8cm}
\vspace*{0cm}
\caption{Broad-band spectrum for four epochs of contemporaneous radio and infrared data, three of which also include X-ray points. Since the source is in a soft state by this time, the radio spectrum (of the first epoch) is relatively steep and optically thin. The infrared points (from Nagata et al. 2003) are in excess of the radio, suggesting that emission from the accretion disc and/or irradiation of the companion star dominate this region of the spectrum. This is in direct contrast to the low/hard state spectra of e.g. Fender (2001) and Markoff et al. (2001) where the jet spectrum dominates.}
\label{spectrum}
\end{center}
\end{figure}

\section{Discussion}

As has been the case for the majority of recent (i.e. observed by {\sl RXTE}) X-ray transient events, the 2003 outburst of XTE J1720$-$318 did not simply display the initial soft spectrum and FRED lightcurve traditionally expected for a ``canonical'' source, but instead showed more complex spectral behaviour. The hard state behaviour at the onset of the outburst adds support to the suggestion of Brocksopp et al. (2002, 2004) that it is this hard state behaviour that holds the key to the outburst mechanism. In particular, since the low/hard state is associated with a jet, the inclusion of the hard spectrum and jet behaviour in outburst models are of great importance. The initial low/hard state period of XTE J1720$-$318 was not monitored in the radio but we were able to obtain radio observations, and detect a radio source, during the later low/hard state period. 

The inset plot in the bottom panel of Fig.~\ref{lightcurve} shows the early radio lightcurve on an expanded time axis. During the first and last epochs shown in this inset panel the 4.9 and 8.5 GHz fluxes are in good agreement with each other, suggesting the presence of at least some partially self-absorbed emission. The intervening epochs of data (obtained at two frequencies simultaneously) show a discrepancy between the two frequencies, indicating that the material responsible for the radio emission had expanded and become optically thin. We list the spectral indices ($\alpha$, where $S_{\nu}\propto\nu^{\alpha}$) in Table~\ref{alpha} and plot them in Fig.~\ref{alpha-plot}. It is clear from Fig.~\ref{alpha-plot} that the radio source becomes significantly optically thin on two separate occasions (MJD 52656, 52659), the optical depth increasing during the intervening epoch, and suggesting that there were actually two ejection events. This is common in X-ray transient events and a sequence of radio ejections have been observed in all sources for which sufficient monitoring has taken place (Brocksopp et al. 2002 and references therein). When the source reappears during the low/hard state it again has a flat spectrum, as we would expect for this state (e.g. Fender 2001).

In Fig.~\ref{spectrum} we plot the broad-band spectrum for the four epochs of contemporaneous radio and infrared data (Nagata et al. 2003); {\sl RXTE} also obtained data-points during three of these epochs. We assume $E(B-V)$=2.43 (T. Nagata, priv. comm.), account for reddening of the infrared data using Cardelli, Clayton \& Mathis (1989) and convert the brightness values to flux densities using Bessel et al. (1998). We further assume that 1 Crab = 75 {\sl RXTE}/ASM counts~sec$^{-1}$ = 1.06 mJy (Levine et al. 1996). By the time of these plotted epochs the X-ray source had made the transition to a soft state and this is reflected in the spectrum. The radio spectrum (MJD 52657) is relatively steep and optically thin (although still partially self-absorbed relative to some of the other epochs in Fig.~\ref{alpha-plot}) whereas the infrared data is showing a clear excess above the radio flux densities. This is very different from the low/hard state spectra in Brocksopp et al. (2004), Fender (2001) and Markoff et al. (2001) where the infrared emission appears to be dominated by the jet synchrotron spectrum; in particular the spectral turn-over between self-absorbed and optically thin synchrotron emission has been detected in the near-infared in GX~339$-$4 (Corbel \& Fender 2002). Instead, the infrared emission of XTE J1720$-$318 during these epochs appears to be dominated by the accretion disc and/or irradiated companion star in agreement with the disc instability model (e.g. Lasota 2001).

The X-ray/radio behaviour over the course of the outburst bears a number of similarities to XTE J1859+226. Following the initial low/hard state, which in both sources lasted $\sim$ days, the radio and soft X-rays reached a peak quasi-simultaneously; in the case of XTE J1859+226 there was a hard X-ray peak {\em prior} to this. Similarly XTE J1720$-$318 reaches a soft X-ray peak approximately coincidentally (or just before) the radio peak; indeed Fender, Belloni \& Gallo (in prep.) suggest that the radio peak and soft X-ray peak are closely related (see also Corbel et al. 2004 who reach a similar conclusion). The steady jet of the initial low-hard state is thought to persist through the period of spectral softening and give way to an optically thin radio ejection just after the soft X-ray peak.

After the initial maximum of XTE J1859+226, it was the hard X-rays that were more closely linked with the radio behaviour. As the X-ray source  decayed from outburst it underwent a series of temporary hardenings, superimposed on a general softening, and each of these hardenings was associated with a new radio ejection (Brocksopp et al. 2002). While the S/N of the hardness ratio is poor, XTE J1720$-$318 also appears to show some level of both intensity and spectral variability during its decay, including at least two glitches. Comparison with other sources such as XTE J1859+226, GRO J1655$-$40 and XTE J1550$-$264 (Brocksopp et al. 2002 and references therein) might suggest that simultaneous radio ejections would have been expected. Given that both glitches and radio ejections are associated with temporary hardenings in these other sources it is tempting to suggest that they are all different manifestations of a jet event. Unfortunately the XTE J1720$-$318 data do not allow us to confirm this, perhaps due to inadequate S/N and/or time resolution. However the data of Nagata et al. (2003) do indeed show simultaneous X-ray/infrared events which coincide with radio non-detections. Again, this may be due to the S/N and time-resolution or it may confirm previous suggestions that some glitches may be events taking place in the accretion disc (e.g. Lasota 2001) and independently of the jet. We also note that, contrary to XTE J1720$-$318, the three sources listed above were in the very high state (or steep power-law state) at the time of the simultaneous glitches/ejections; it is probable that the spectral state of the accretion disc is significant in determining whether or not ejections take place (e.g. Fender, Belloni \& Gallo in prep.).

\section{Conclusions}

We have presented radio observations of the 2003 outburst of the X-ray transient XTE J1720$-$318. The radio source was unresolved and reached a peak of $\sim 5$ mJy. Study of the spectral index showed that, while the source was optically thin throughout, there was some significant variability which we interpret as partially self-absorbed emission at the onset of two discrete ejection events. Following a period of non-detection, the radio source switched on again contemporaneously with the transition to the low/hard state. The broadband spectrum showed that the infrared emission was significantly brighter than the radio synchrotron spectrum; we suggest that the infrared emission was dominated by the accretion disc and/or irradiation of the companion star as expected for a ``soft'' X-ray transient event. The variability during the decay was reminiscent of that of XTE J1859+226; in the case of XTE J1859+226 this variability was interpreted as a sequence of jet ejections. This emphasizes the need for high S/N X-ray hardness and radio monitoring during the ``glitches'' which are often observed superimposed on the decay of the X-ray source.

\section*{acknowledgments}
We are very grateful to Jean Swank, who kindly shared some {\it RXTE}/PCA results with us prior to publication. We are also grateful for the quick-look results provided by the {\it RXTE}/ASM team. The Australia Telescope is funded by the Commonwealth of Australia for operation as a National Facility managed by CSIRO. DCH is a Fellow of the Finnish Academy.

\end{document}